%% file: main.tex
\documentclass[sigconf]{aamas}  % do not change this line!

\pdfoutput=1

\usepackage{booktabs}
\usepackage{gensymb}
\usepackage{multirow}
\usepackage{changepage}

\usepackage{flushend}
\setcopyright{ifaamas}  % do not change this line!
\acmISBN{}  % do not change this line!
\acmConference[AAMAS'20]{Proc.\@ of the 19th International Conference on Autonomous Agents and Multiagent Systems (AAMAS 2020), B.~An, N.~Yorke-Smith, A.~El~Fallah~Seghrouchni, G.~Sukthankar (eds.)}{May 2020}{Auckland, New Zealand}  % do not change this line!
\acmYear{2020}  % do not change this line!
\copyrightyear{2020}  % do not change this line!
\acmPrice{}  % do not change this line!

\begin{document}

\title[Social Diversity and Social Preferences in Mixed-Motive Reinforcement Learning]{Social Diversity and Social Preferences\\in Mixed-Motive Reinforcement Learning}  % put your title here!
% \titlenote{Produces the permission block, and copyright information}

% AAMAS: as appropriate, uncomment one subtitle line; check the CFP
%\subtitle{Extended Abstract}
%\subtitle{Blue Sky Ideas Track}
%\subtitle{JAAMAS Track}
%\subtitle{Demonstration}
%\subtitle{Doctoral Consortium}

% AAMAS: submissions are anonymous for most tracks
% \author{Paper \#1157}  % put your paper number here!

%% example of author block for camera ready version of accepted papers: don't use for anonymous submissions

\author{Kevin R. McKee, Ian Gemp, Brian McWilliams, \\Edgar A. Du\'e\~nez-Guzm\'an, Edward Hughes, \& Joel Z. Leibo}
\affiliation{%
 \institution{DeepMind}
 \city{London}
}
\email{{kevinrmckee, imgemp, bmcw, duenez, edwardhughes, jzl}@google.com}
%% The example's default list of authors is too long for headers
\renewcommand{\shortauthors}{K. McKee et al.}

\begin{abstract}  % put your abstract here!
Recent research on reinforcement learning in pure-conflict and pure-common interest games has emphasized the importance of population heterogeneity. In contrast, studies of reinforcement learning in mixed-motive games have primarily leveraged homogeneous approaches. Given the defining characteristic of mixed-motive games---the imperfect correlation of incentives between group members---we study the effect of population heterogeneity on mixed-motive reinforcement learning. We draw on \textit{interdependence theory} from social psychology and imbue reinforcement learning agents with Social Value Orientation (SVO), a flexible formalization of preferences over group outcome distributions. We subsequently explore the effects of diversity in SVO on populations of reinforcement learning agents in two mixed-motive Markov games. We demonstrate that heterogeneity in SVO generates meaningful and complex behavioral variation among agents similar to that suggested by interdependence theory. Empirical results in these mixed-motive dilemmas suggest agents trained in heterogeneous populations develop particularly generalized, high-performing policies relative to those trained in homogeneous populations.
\end{abstract}

  % put your semicolon-separated keywords here!

\maketitle

%%%%%%%%%%%%%%%%%%%%%%%%%%%%%%%%%%%%%%%%%%%%%%%%%%%%%%%%%%%%%%%%%%%%%%%%%%%%%%%%%%%%%%%%%%%%%%%%%%%%%%%%%
%% start of main body of paper

\input{body}

%%%%%%%%%%%%%%%%%%%%%%%%%%%%%%%%%%%%%%%%%%%%%%%%%%%%%%%%%%%%%%%%%%%%%%%%%%%%%%%%%%%%%%%%%%%%%%%%%%%%%%%%%
%% bibliography: see CFP for number of permitted pages

% \bibliographystyle{ACM-Reference-Format}  % do not change this line!
\bibliographystyle{plain}
\bibliography{main.bib}  % put name of your .bib file here

\end{document}

%% file: body.tex
\section{Introduction}

In multi-agent reinforcement learning, the actions of one agent can influence the experience and outcomes for other agents---that is, agents are \textit{interdependent}. Interdependent interactions can be sorted into two categories based on the alignment of incentives for the agents involved \cite{schelling1960strategy}:

\begin{enumerate}
    \item \textit{Pure-motive} interactions, in which the group's incentives are either entirely aligned (pure-common interest) or entirely opposed (pure-conflict),
    \item and \textit{mixed-motive} interactions, in which the group's incentives are sometimes aligned and sometimes in conflict.\footnote{When Schelling originally introduced the pure- and mixed-motive framework, he explained, ``Mixed-motive refers not, of course, to an individual's lack of clarity about his own preferences but rather to the ambivalence of his relation to the other player---the mixture of mutual dependence and conflict, of partnership and competition'' \cite{schelling1960strategy}.}
\end{enumerate}

\noindent Examples of the former include games such as Hanabi \cite{bard2019hanabi} and Go \cite{silver2018general}. The latter category is typified by games like the Prisoner's Dilemma \cite{tucker1950two, sandholm1996multiagent} and the tragedy of the commons \cite{hardin1968tragedy, leibo2017multi}. This categorical distinction is especially relevant for spatially and temporally extended Markov games. In these games, interdependence emerges both as a direct impact of one agent's actions on another's outcomes and as an indirect effect of each agent on the state of the substrate environment in which others co-exist.

\begin{figure}[!b]
    \centering
    \includegraphics[width=8.125cm]{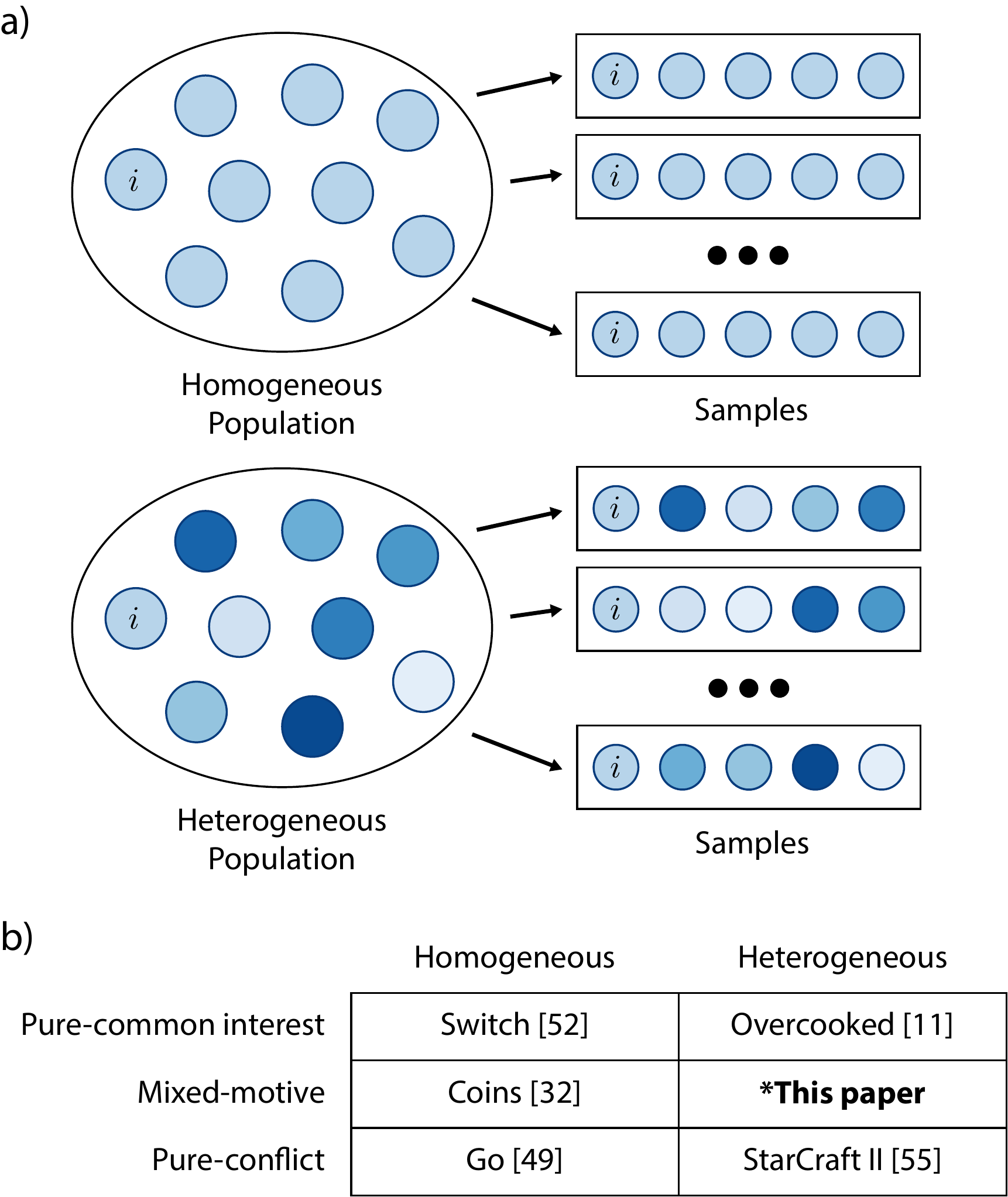}
    \caption{Homogeneity and heterogeneity in population-based multi-agent reinforcement learning. (a) Population homogeneity and heterogeneity result in different training experiences for a given agent \textit{\textbf{i}}. In the homogeneous case, agent policies are either identical or very similar (e.g., due to identical training distributions or shared motivations). In the heterogeneous setting, a given agent \textit{\textbf{i}} encounters a range of group compositions over time. The variability in policies can stem from agents training under different distributions or with different motivations. (b) Representative examples of previous multi-agent reinforcement learning research. We study the mixed-motive, heterogeneous setting.}
    \label{fig:geneity}
\end{figure}

In pure-conflict reinforcement learning, self-play solutions for Markov games \cite{bansal2017emergent, heinrich2016deep, silver2018general} have gradually given way to population-based approaches \cite{jaderberg2019human, vinyals2019grandmaster} (Figure \ref{fig:geneity}a). A central impetus for this shift has been an interest in ensuring agent performance is robust to opponent heterogeneity (i.e., variation in the set of potential opponent policies). Similarly, recent work on pure-common interest reinforcement learning in Markov games has highlighted the importance of robustness to diverse partner policies \cite{amato2013decentralized, carroll2019utility}. In both of these contexts, it is desirable to train agents capable of adapting and best responding to a wide range of potential policy sets.

In mixed-motive Markov games, the effects of partner heterogeneity have not received much attention. Most mixed-motive reinforcement learning research has produced policies through self-play \cite{lerer2017maintaining, peysakhovich2017consequentialist} or co-training policies in fixed groups \cite{leibo2017multi, hughes2018inequity}. Such methods foster homogeneity in the set of policies each agent encounters.

We aim to introduce policy heterogeneity into mixed-motive reinforcement learning (Figure \ref{fig:geneity}b). In recent years, a growing body of work has explored the effects of furnishing agents in mixed-motive games with various motivations such as inequity aversion \cite{hughes2018inequity, wang2019evolving}, social imitation \cite{eccles2019imitation}, and social influence \cite{jaques2019social}. Thus, a natural starting point for the study of heterogeneity is to explore the effects of diversity in intrinsic motivation \cite{singh2004intrinsically}. Here we endow our agents with Social Value Orientation (SVO), an intrinsic motivation to prefer certain group reward distributions between self and others.

\begin{figure}[b]
    \centering
    \includegraphics[width=7.5cm]{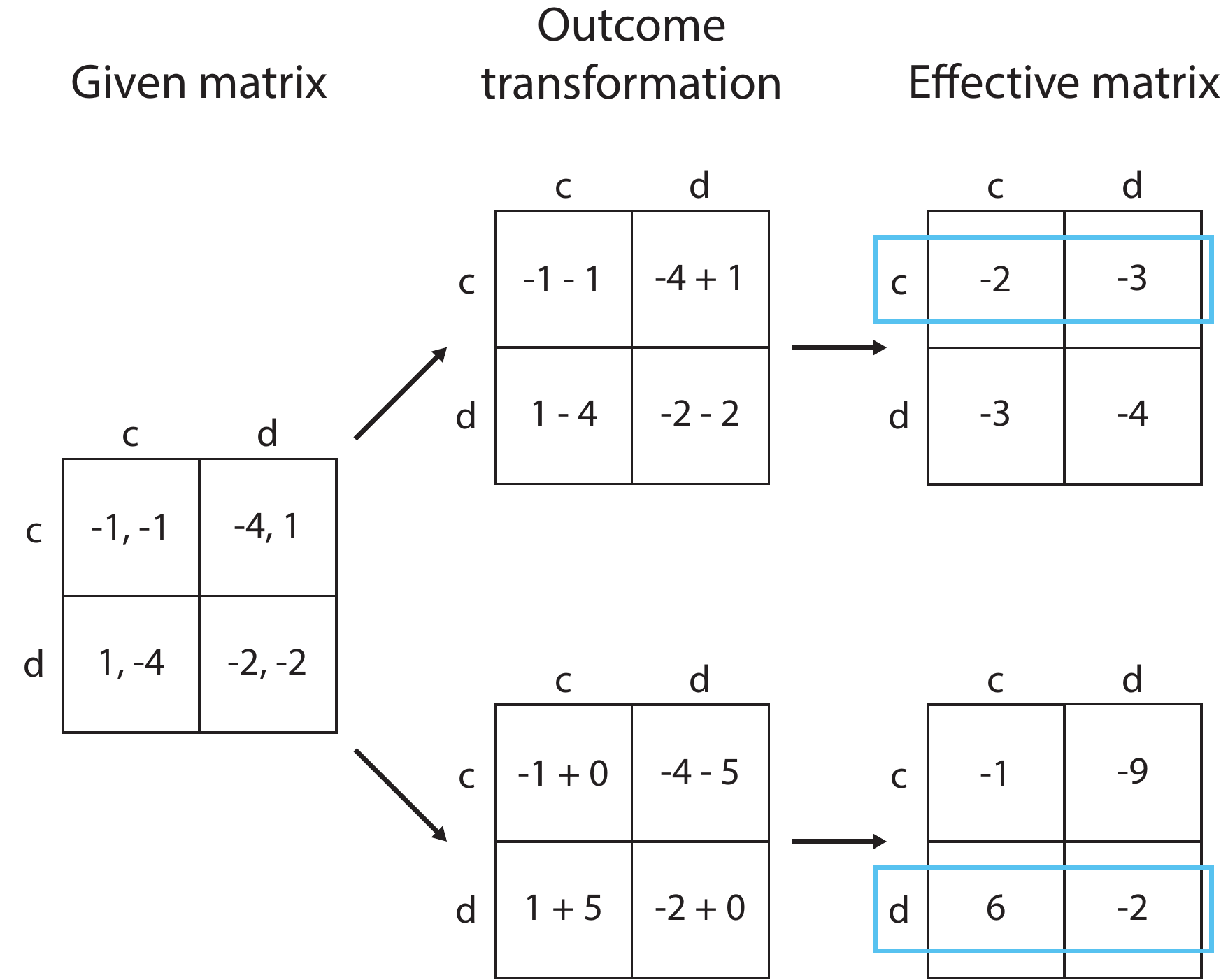}
    \caption{Interdependence theory. The top pathway depicts a transformation process for a row player who has altruistic preferences. In this case, the outcome transformation directly transfers the column player's payoff into the effective matrix. The bottom pathway shows a transformation process for a row player with competitive preferences, who finds it rewarding to maximize the distance between their payoff and the column player's payoff. These two outcome transformations suggest different dominant strategies (highlighted in blue).}
    \label{fig:outcome_transformation}
\end{figure}

Psychology and economics research has repeatedly demonstrated that human groups sustain high levels of cooperation across different games through heterogeneous distributive preferences \cite{batson2012history, cooper2016other, eckel1996altruism, rushton1981altruistic, simon1993altruism}. A particularly compelling account from \textit{interdependence theory} holds that humans deviate from game theoretic predictions in economic games because each player acts not on the ``given matrix'' of a game---which reflects the extrinsic payoffs set forth by the game rules---but on an ``effective matrix'', which represents the set of outcomes as subjectively evaluated by that player \cite{kelley1978interpersonal}. Players receive the given matrix from their environment and subsequently apply an ``outcome transformation'' reflecting their individual preferences for various outcome distributions. The combination of the given matrix and the outcome transformation form the effective matrix (Figure \ref{fig:outcome_transformation}). Though an individual's social preferences may decrease their given payoff in a single game, groups with diverse sets of preferences are capable of resisting suboptimal Nash equilibria.

In multi-agent reinforcement learning, reward sharing is commonly used to resolve mixed-motive dilemmas \cite{hughes2018inequity, peysakhovich2018prosocial, sunehag2018value, wang2019evolving}. To date, agent hyperparameters controlling reward mixture have typically been shared. This approach implies a homogeneous population of policies, echoing the representative agent assumption from economics \cite{hartley1996retrospectives, kirman1992whom}. The continued reliance on shared reward mixtures is especially striking considering that the ability to capture heterogeneity is a key strength of agent-based models over other modeling approaches \cite{haldane2019drawing}.

Homogeneous populations often fall prey to a peculiar variant of the lazy agent problem \cite{sunehag2018value}, wherein one or more agents begin ignoring the individual learning task at hand and instead optimize for the shared reward \cite{hughes2018inequity}. These shared-reward agents shoulder the burden of prosocial work, in a manner invariant to radical shifts in group composition across episodes. This ``specialization'' represents a failure of training to generate generalized policies.

To investigate the effects of heterogeneity in mixed-motive reinforcement learning, we introduce a novel, generalized mechanism for reward sharing. We derive this reward-sharing mechanism, Social Value Orientation (SVO), from studies of human cooperative behavior in social psychology. We show that across several games, heterogeneous distributions of these social preferences within groups generate more generalized individual policies than do homogeneous distributions. We subsequently explore how heterogeneity in SVO sustains positive group outcomes. In doing so, we demonstrate that this formalization of social preferences leads agents to discover specific prosocial behaviors relevant to each environment.

\section{Agents}

\subsection{Multi-agent reinforcement learning and Markov games}

In this work, we consider $n$-player partially observable Markov games. A partially observable Markov game $\mathcal{M}$ is defined on a finite set of states $\mathcal{S}$. The game is endowed with an observation function $O: \mathcal{S} \times \{1,\dots,n\} \rightarrow \mathbb{R}^d$; a set of available actions for each player, $\mathcal{A}_1,\dots,\mathcal{A}_n$; and a stochastic transition function $\mathcal{T}: \mathcal{S} \times \mathcal{A}_1 \times \cdots \times \mathcal{A}_n \rightarrow \Delta(\mathcal{S})$, which maps from the joint actions taken by the $n$ players to the set of discrete probability distributions over states. From each state, players take a joint action $\vec{a} =(a_1,\dots,a_n) \in \mathcal{A}_1,\dots,\mathcal{A}_n$.

Each agent $i$ independently experiences the environment and learns a behavior policy $\pi(a_i|o_i)$ based on its own observation $o_i = O(s,i)$ and (scalar) extrinsic reward $r_i(s,\vec{a})$. Each agent learns a policy which maximizes a long term $\gamma$-discounted payoff defined as:

\begin{equation}
    V_{\vec{\pi}_i}(s_0) = \mathbb{E} \left[ \sum \limits_{t=0}^{\infty} \gamma^t U_i(s_t, \vec{o}_t, \vec{a}_t) | \vec{a}_t \sim \vec{\pi}_t, s_{t+1} \sim \mathcal{T}(s_t, \vec{a}_t) \right]
    \label{eq:max_v}
\end{equation}

\noindent where $U_i(s_t, \vec{o}_t, \vec{a}_t)$ is a utility function and $\vec{o} = (o_1,\dots, o_n)$ for simplicity. In standard reinforcement learning, the utility function maps directly to the extrinsic reward provided by the environment.

\subsection{Social Value Orientation}

Here we introduce Social Value Orientation (SVO), an intrinsic motivation to prefer certain group reward distributions between self and others.

\begin{figure}
    \centering
    \includegraphics[width=7.5cm]{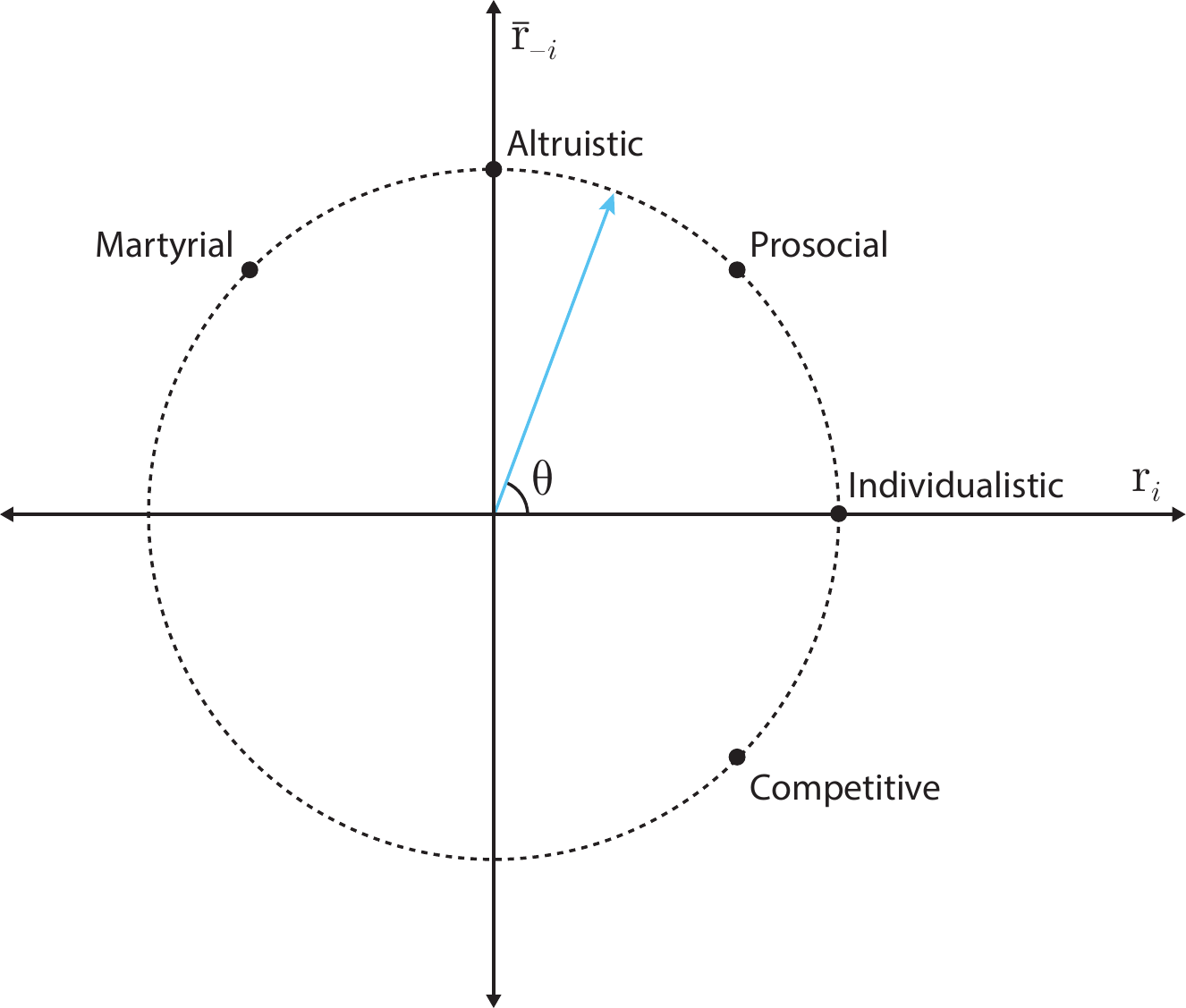}
    \caption{Reward angles and the ring formulation of Social Value Orientation (SVO). Reward angles are a scalar representation of the tradeoff between an agent's own reward and the reward of other agents in the environment. The reward angle an agent prefers is its SVO.}
    \label{fig:svo_ring}
\end{figure}

We introduce the concept of a \emph{reward angle} as a scalar representation of the observed distribution of reward between player $i$ and all other players in the group. The size of the angle formed by these two scalars represents the relative distribution of reward between self and others (Figure \ref{fig:svo_ring}). Given a group of size $n$, its corresponding reward vector is $\vec{R} = (r_1,\ldots, r_n)$. The reward angle for player $i$ is:

\begin{equation}
    \theta(\vec{R}) = \mathrm{atan}\left(\frac{\bar{r}_{-i}}{r_{i}}\right)
\end{equation}

\noindent where $\bar{r}_{-i}$ is a statistic summarizing the rewards of all other group members. We choose the arithmetic mean $\bar{r}_{-i} = \frac{1}{{n - 1}} \sum_{j \neq i} r_j$ \cite{nisbett1985perception}. Note that reward angles are invariant to the scalar magnitude of $\vec{R}$.

The Social Value Orientation, $\theta^{\text{SVO}}$, for player $i$ is player $i$'s target distribution of reward among group members. We use the difference between the observed reward angle and the targeted SVO to calculate an intrinsic reward. Combining this intrinsic reward with the extrinsic reward signal from the environment, we can define the following utility function $U_i$ to be maximized in Eq.~\eqref{eq:max_v}:

\begin{equation}
    U_{i}(s, o_i, a_i) = r_{i} - w \cdot \lvert \theta^\text{SVO} - \theta(\vec{R}) \rvert
    \label{eqn:utility}
\end{equation}

\noindent where $w$ is a weight term controlling the effect of Social Value Orientation on $U_{i}$.

In constructing this utility function, we follow a standard approach in mixed-motive reinforcement learning research \cite{hughes2018inequity, jaques2019social, wang2019evolving} which provides agents with an overall reward signal combining extrinsic and intrinsic reward \cite{singh2004intrinsically}. This approach parallels interdependence theory, wherein the effective matrix is formed from the combination of the given matrix and the outcome transformation, and ultimately serves as the basis of actors' decisions \cite{kelley1978interpersonal}.

For the exploratory work we detail in the following sections, we restrict our experiments to SVO in the non-negative quadrant (all $\theta \in [0\degree, 90\degree]$). The preference profiles in the non-negative quadrant provide the closest match to parameterizations in previous multi-agent research on reward sharing. Nonetheless, we note that interesting preference profiles exist throughout the entire ring \cite{murphy2014social}.

\subsection{Algorithm}

We deploy advantage actor-critic (A2C) as the learning algorithm for our agents \cite{mnih2016asynchronous}. A2C maintains both value (critic) and policy (actor) estimates using a deep neural network. The policy is updated according to the REINFORCE policy-gradient method, using a value estimate as a baseline to reduce variance. Our neural network comprises a convolutional layer, a feedforward module, an LSTM with contrastive predictive coding \cite{oord2018representation}, and linear readouts for policy and value. We apply temporal smoothing to observed rewards within the model's intrinsic motivation function, as described by \cite{hughes2018inequity}.

We use a distributed, asynchronous framework for training \cite{wang2019evolving}. We train populations of $N = 30$ agents with policies $\{\pi_i\}$. For each population, we sample $n = 5$ players at a time to populate each of 100 arenas running in parallel (see also Figure \ref{fig:geneity}a, in which arenas are represented as ``samples'' from the agent population). Each arena is an instantiation of a single episode of the environment. Within each arena, the sampled agents play an episode of the environment, after which a new group is sampled. Episode trajectories last 1000 steps and are written to queues for learning. Weights are updated from queues using V-Trace \cite{espeholt2018impala}.

\section{Mixed-motive games}
\subsection{Intertemporal social dilemmas}

\begin{figure*}[t]
    \centering
    \includegraphics[width=17.5cm]{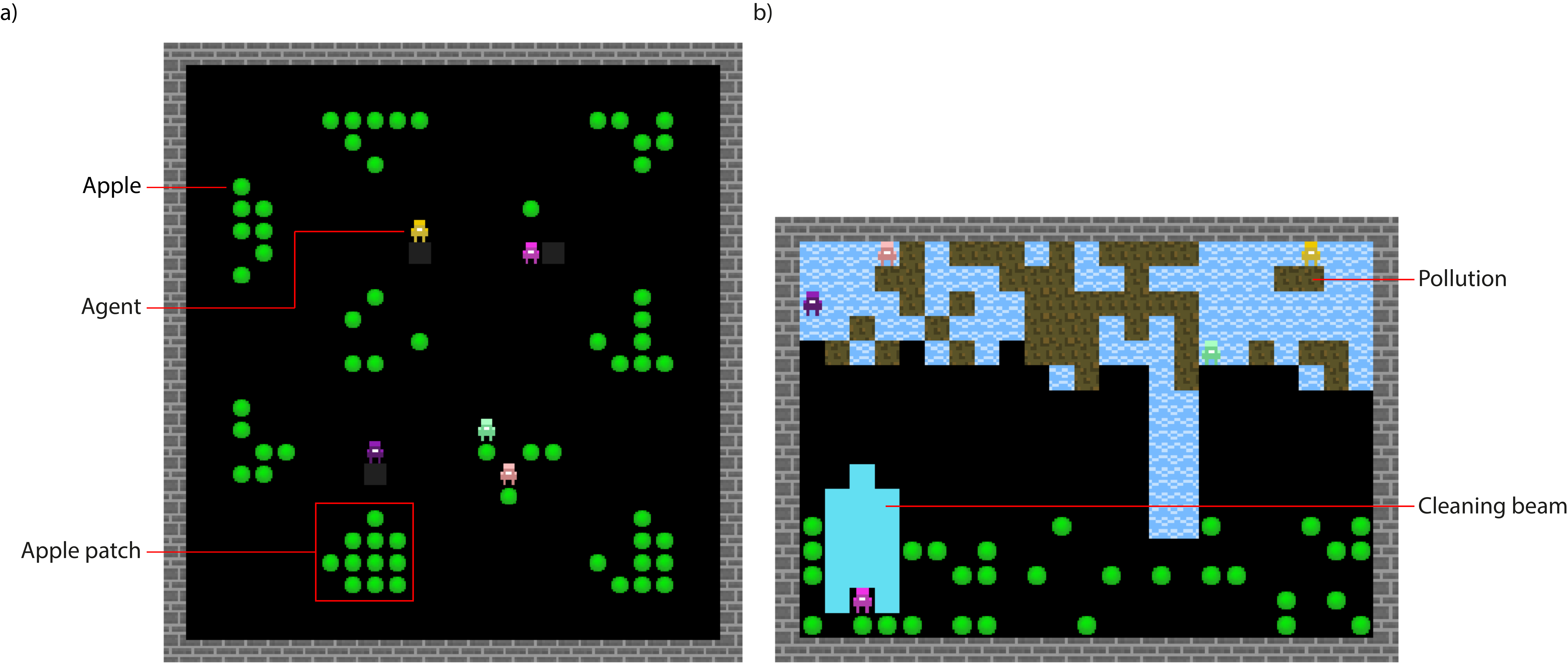}
    \caption{Screenshots of gameplay from (a) HarvestPatch and (b) Cleanup.}
    \label{fig:gameplay_screenshot}
\end{figure*}

For our experiments, we consider two temporally and spatially extended mixed-motive games played with group size $n = 5$: HarvestPatch and Cleanup. These two environments are intertemporal social dilemmas, a particular class of mixed-motive Markov games (c.f., \cite{littman1994markov}).

Intertemporal social dilemmas are group situations which present a tension between short-term individual incentives and the long-term collective interest \cite{hughes2018inequity}. Each individual has the option of behaving prosocially (cooperation) or selfishly (defection). Though unanimous cooperation generates welfare-maximizing outcomes in the long term, on short timescales the personal benefits of acting selfishly strictly dominate those of prosocial behavior. Thus, though all members of the group prefer the rewards of mutual cooperation, the intertemporal incentive structure pushes groups toward welfare-suppressing equilibria. Previous work has evaluated the game theoretic properties of intertemporal social dilemmas \cite{hughes2018inequity}.

\subsection{HarvestPatch}

HarvestPatch is a variant of the common-pool resource appropriation game Harvest \cite{hughes2018inequity} (Figure \ref{fig:gameplay_screenshot}a). Players are rewarded for collecting apples (reward $+1$) within a $24 \times 26$ gridworld environment. Apples regrow after being harvested at a rate dependent on the number of unharvested apples within a regrowth radius of 3. If there are no apples within its radius, an apple cannot regrow. At the beginning of each episode, apples are probabilistically spawned in a hex-like pattern of patches, such that each apple is within the regrowth radius of all other apples in its patch and outside of the regrowth radius of apples in all other patches. This creates localized stock and flow properties \cite{gardner1990nature} for each apple patch. Each patch is irreversibly depleted when all of its apples have been harvested---regardless of how many apples remain in other patches. Players are also able to use a beam to punish other players (reward $-50$), at a small cost to themselves (reward $-1$). This enables the possible use of punishment to discourage free-riding \cite{henrich2006cooperation, o2008constraining}.

A group can achieve indefinite sustainable harvesting by abstaining from eating ``endangered apples'' (apples which are the last unharvested apple remaining in their patch). However, the reward for sustainable harvesting only manifests after a period of regrowth if all players abstain. In contrast, an individual is immediately and unilaterally guaranteed the reward for eating an endangered apple if it acts greedily. This creates a dilemma juxtaposing the short-term individual temptation to maximize reward through unsustainable behavior and the long-term group interest of generating higher reward by acting sustainably.

In HarvestPatch, episodes last 1000 steps. Each agent's observability is limited to a $15 \times 15$ RGB window, centered on its current location. The action space consists of movement, rotation, and use of the punishment beam (8 actions total).

\subsection{Cleanup}

Cleanup \cite{hughes2018inequity} is a public goods game (Figure \ref{fig:gameplay_screenshot}b). Players are again rewarded for collecting apples (reward $+1$) within a $25 \times 18$ gridworld environment. In Cleanup, apples grow in an orchard at a rate inversely related to the cleanliness of a nearby river. The river accumulates pollution with a constant probability over time. Beyond a certain threshold of pollution, the apple growth rate in the orchard drops to zero. Players have an additional action allowing them to clean a small amount of pollution from the river. However, the cleaning action only works on pollution within a small distance in front of the agents, requiring them to physically leave the apple orchard to clean the river. Thus, players maintain the public good of orchard regrowth through effortful contributions. As in HarvestPatch, players are able to use a beam to punish other players (reward $-50$), at a small cost to themselves (reward $-1$).

A group can achieve continuous apple growth in the orchard by keeping the pollution levels of the river consistently low over time. However, on short timescales, each player would prefer to collect apples in the orchard while other players provide the public good in the river. This creates a tension between the short-term individual incentive to maximize reward by staying in the orchard and the long-term group interest of maintaining the public good through sustained contributions over time.

Episodes last 1000 steps. Agent observability is again limited to a $15\times15$ RGB window, centered on the agent's current location. In Cleanup, agents have an additional action for cleaning (9 actions total).

\section{Results}

\subsection{Social diversity and agent generality}

\begin{figure}[t]
    \centering
    \includegraphics[width=8.5cm]{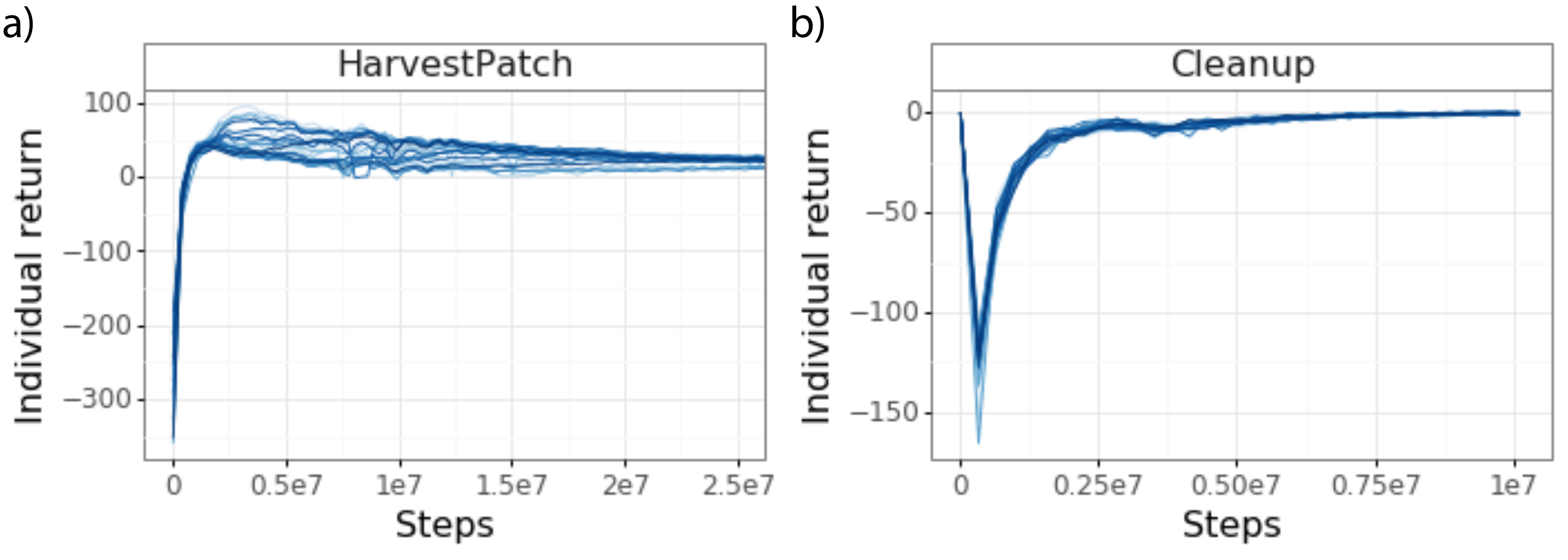}
    \caption{Episode rewards for homogeneous selfish populations playing (a) HarvestPatch and (b) Cleanup. Each individual line shows a single agent's return over training.}
    \label{fig:agent_training}
\end{figure}

We began by training 12 homogeneous populations per task, with $N = 30$: four consisting of individualistic agents (all $\theta = 0\degree$), four of prosocial agents (all $\theta = 45\degree$), and four of altruistic agents (all $\theta = 90\degree$) agents. These resemble previous approaches using selfishness \cite{perolat2017multi}, inequity aversion \cite{hughes2018inequity, wang2019evolving}, and strong reward sharing \cite{peysakhovich2018prosocial, sunehag2018value}, respectively. The population training curves for homogeneous selfish populations closely resembled group training curves from previous studies \cite{perolat2017multi, hughes2018inequity} (see sample population training curves in Figure \ref{fig:agent_training}). In particular, performance in both environments generated negative returns at the beginning of training due to high-frequency use of the punishment beam. Agents quickly improved performance by learning not to punish one another, but failed to learn cooperative policies. Ultimately, selfish agents were unable to consistently avoid the tragedy of the commons in HarvestPatch or provide public goods in Cleanup. 

\begin{figure}[b]
    \centering
    \includegraphics[width=8.5cm]{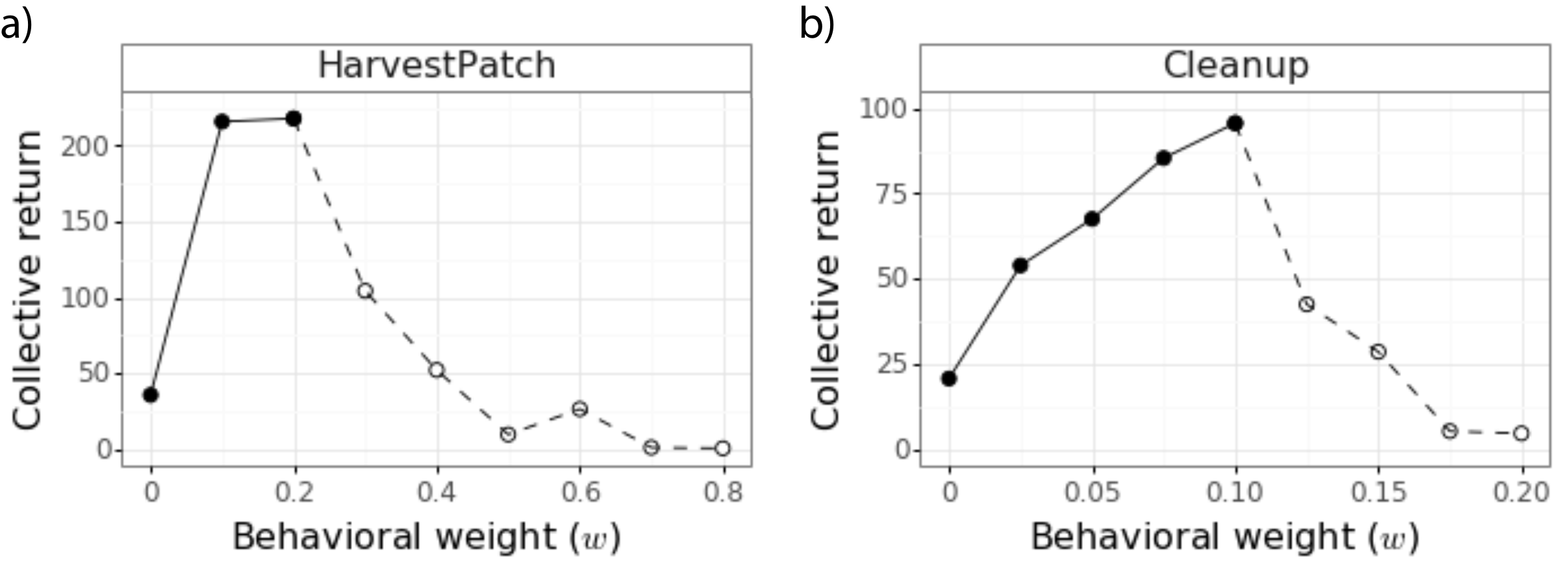}
    \caption{Equilibrium collective return for homogeneous populations of altruistic agents in (a) HarvestPatch and (b) Cleanup. Closed dots reflect populations in which all agents receive positive returns at equilibrium. Open dots indicate populations in which some agents receive zero or negative reward.}
    \label{fig:agent_behavioral_weight}
\end{figure}

Optimal hyperparameter values may vary between HarvestPatch and Cleanup. Thus, we selected the weight values $w$ for the two tasks by conducting an initial sweep over $w$ with homogeneous populations of $N = 20$ altruistic agents (all $\theta = 90\degree$). In HarvestPatch, a weight $w = 0.2$ produced the highest collective returns across several runs (Figure \ref{fig:agent_behavioral_weight}a). In Cleanup, a weight $w = 0.1$ produced the highest collective returns across several runs (Figure \ref{fig:agent_behavioral_weight}b).

As expected, in HarvestPatch, the highest collective returns among the homogeneous populations were achieved by the altruistic populations (Table \ref{tab:homogeneity_heterogeneity_collective_return}, Homogeneous row). The prosocial and individualistic populations performed substantially worse. In Cleanup, the highest collective returns similarly emerged among the altruistic populations. The populations of prosocial and individualistic agents, in contrast, achieved near-zero collective returns.

\begin{table}[t]
    \begin{tabular}{ |c|c|c|c|}
        \hline
         & Mean SVO & HarvestPatch & Cleanup \\ 
        \hline
        \multirow{3}{6em}{Homogeneous} & $0\degree$ & 587.6 (101.7) & -9.9 (11.7) \\ 
        & $45\degree$ & 665.9 (52.4) & 1.1 (2.1) \\ 
        & $90\degree$ & \textbf{1552.7 (248.2)} & 563.8 (235.2) \\
        \hline
        \multirow{5}{6em}{Heterogeneous} & $15\degree$ & 553.4 (574.6) & -0.1 (5.7)\\ 
        & $30\degree$ & 658.7 (107.1) & 2.0 (2.4) \\ 
        & $45\degree$ & 764.1 (236.3) & 6.3 (7.1) \\ 
        & $60\degree$ & 860.9 (121.5) & 318.5 (335.0) \\ 
        & $75\degree$ & 1167.9 (232.6) & \textbf{1938.5 (560.6)} \\ 
        \hline
    \end{tabular} \medskip
    \caption{Mean collective returns achieved at equilibrium by homogeneous and heterogeneous populations. Standard deviations are reported in parentheses.}
    \label{tab:homogeneity_heterogeneity_collective_return}
\end{table}

We next trained 80 heterogeneous populations per task. To generate each heterogeneous population, we sampled $N = 30$ SVO values from a normal distribution with a specified mean and dispersion. Since we treated SVO as a bounded variable for these initial experiments, we selected five equally spaced values from $15\degree$ to $75\degree$ to act as population means and and four equally spaced values from $5.6\degree$ ($\frac{\pi}{32}$ radians) to $11.3\degree$  ($\frac{\pi}{16}$ radians) to act as population standard deviations. For each mean-standard deviation pair, we generated four populations using different random seeds. We used the same weights $w$ as for the homogeneous populations.

\begin{table}[b]
    \begin{tabular}{ |c|c|c|c|}
        \hline
         & Mean SVO & HarvestPatch & Cleanup \\ 
        \hline
        \multirow{3}{6em}{Homogeneous} & $0\degree$ & 0.90 (0.09) & 0.16 (0.09) \\ 
        & $45\degree$ & \textbf{0.97 (0.01)} & 0.54 (0.07) \\ 
        & $90\degree$ & 0.29 (0.03) & 0.41 (0.08) \\
        \hline
        \multirow{5}{6em}{Heterogeneous} & $15\degree$ & 0.90 (0.13) & 0.34 (0.09)\\ 
        & $30\degree$ & 0.94 (0.06) & 0.40 (0.09) \\ 
        & $45\degree$ & 0.95 (0.03) & 0.38 (0.10) \\ 
        & $60\degree$ & 0.91 (0.02) & 0.64 (0.21) \\ 
        & $75\degree$ & 0.76 (0.04) & \textbf{0.87 (0.08)} \\ 
        \hline
    \end{tabular} \medskip
    \caption{Mean equality scores achieved at equilibrium by homogeneous and heterogeneous populations. Equality is calculated as the inverse Gini coefficient, with 0 representing all reward being received by a single agent and 1 representing all agents receiving an identical positive reward. Standard deviations are reported in parentheses.}
    \label{tab:homogeneity_heterogeneity_equality}
\end{table}

Among the heterogeneous populations, we observed the highest equilibrium collective returns among the $75\degree$ population (Table \ref{tab:homogeneity_heterogeneity_collective_return}, Heterogeneous row). In HarvestPatch, the performance of homogeneous altruistic populations outstripped the performance of the $75\degree$ populations. In Cleanup, the reverse pattern emerged: the highest collective returns among all populations are achieved by the heterogeneous $75\degree$ populations.

We unexpectedly find that homogeneous populations of altruistic agents produced lower equality scores than most other homogeneous and heterogeneous populations (Table \ref{tab:homogeneity_heterogeneity_equality}). Homogeneous, altruistic populations earned relatively high collective returns in both tasks. However, in each case the produced rewards were concentrated in a small proportion of the population. Agents in these homogeneous populations appear to adopt a lazy-agent approach \cite{sunehag2018value} to resolve the conflict between the group's shared preferences for selfless reward distributions. To break the symmetry of this dilemma, most agents in the population selflessly support collective action, thereby optimizing for their social preferences. A smaller number of agents then specialize in accepting the generated reward---shouldering the ``burden'' of being selfish, in contravention of their intrinsic preferences. This result highlights a drawback of using collective return as a performance metric. Though collective return is the traditional social outcome metric used in multi-agent reinforcement learning, it can mask high levels of inequality.

We therefore revisited population performance by measuring median return, which incorporates signal concerning the efficiency \textit{and} the equality of a group's outcome distribution \cite{blakely2001difference}. Median return can help estimate the generality of learned policies within homogeneous and heterogeneous populations. We compare median return for the two population types by measuring the median return for each population after it reaches equilibrium. We conduct a Welch's \textit{t}-test and report the resulting \textit{t}-statistic, degrees of freedom, and \textit{p}-value. We subsequently provide effect estimates ($\beta$) and \textit{p}-values from linear models regressing median return on the mean and standard deviation of population SVO.

\begin{figure}
    \centering
    \includegraphics[width=8.5cm]{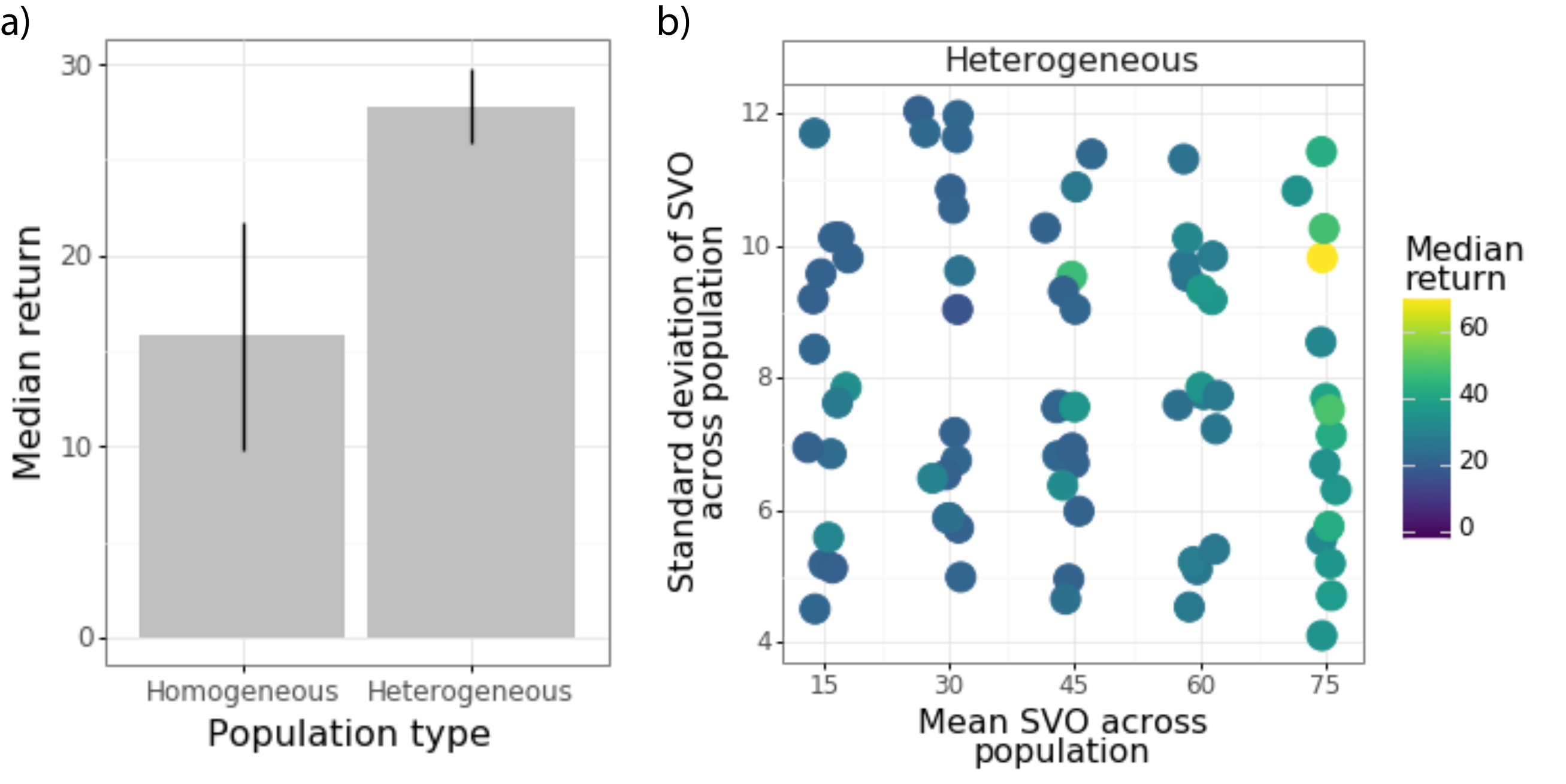}
    \caption{Equilibrium performance of homogeneous and heterogeneous agent populations in HarvestPatch. (a) Heterogeneous populations enjoyed significantly higher median return at equilibrium. (b) Among heterogeneous populations, the highest median returns emerged among the populations whose SVO distributions have both a high mean and a high standard deviation.}
    \label{fig:patch_training_comparison}
\end{figure}

\begin{figure}
    \centering
    \includegraphics[width=8.5cm]{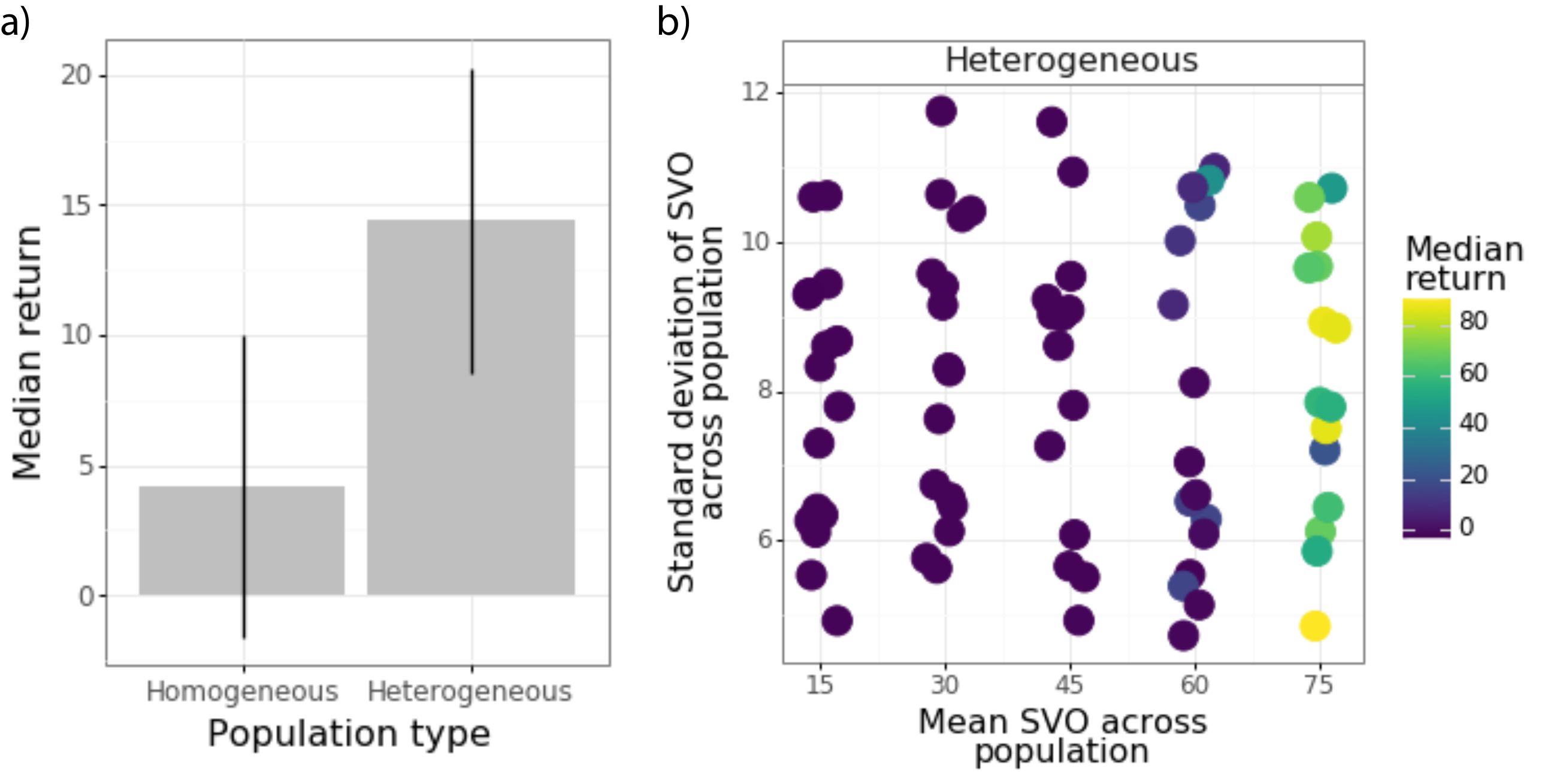}
    \caption{Equilibrium performance of homogeneous and heterogeneous agent populations in Cleanup. (a) Heterogeneous populations received higher median return at equilibrium. (b) Among heterogeneous populations, the highest median returns emerged among the populations with a high mean SVO.}
    \label{fig:cleanup_training_comparison}
\end{figure}

Figures \ref{fig:patch_training_comparison} and \ref{fig:cleanup_training_comparison} show the generality of policies trained in HarvestPatch and Cleanup, respectively. In HarvestPatch, heterogeneous populations enjoyed significantly higher median return ($\mu = 27.8$) than homogeneous populations ($\mu = 15.8$) at equilibrium, $t(13.7) = 4.21, p < 0.001$ (Figure \ref{fig:patch_training_comparison}a). Among heterogeneous populations, a clear pattern could be observed: the higher the population mean SVO, the higher the median return received (Figure \ref{fig:patch_training_comparison}b). Specifically for populations with high mean SVO, median return appeared to increase slightly when the SVO distribution was more dispersed. When tested with a linear model regressing median return on the mean and standard deviation of SVO, these trends primarily manifested as an interaction effect between mean population SVO and the standard deviation of population SVO, $\beta = 0.025$, $p = 0.030$. In Cleanup, heterogeneous populations received significantly higher median return ($\mu = 14.4$) than homogeneous populations ($\mu = 4.2$) at equilibrium, $t(35.9) = 2.43, p = 0.020$ (Figure \ref{fig:cleanup_training_comparison}a). Among heterogeneous populations, the highest median returns were observed in tandem with high mean SVO (Figure \ref{fig:cleanup_training_comparison}b). However, in this case the effect of the interaction between mean SVO and standard deviation of SVO was non-significant, $p = 0.30$.

In summary, our comparison of homogeneous and heterogeneous populations shows that populations of altruists performed effectively in traditional terms (collective return). However, these populations produced highly specialized agents, resulting in undesirably low equality metrics. Populations with diverse SVO distributions were able to circumvent this symmetry-breaking problem and achieve high levels of median return in HarvestPatch and Cleanup.

\subsection{Social preferences and prosocial behavior}

\begin{figure}[b]
    \centering
    \includegraphics[height=4.5cm]{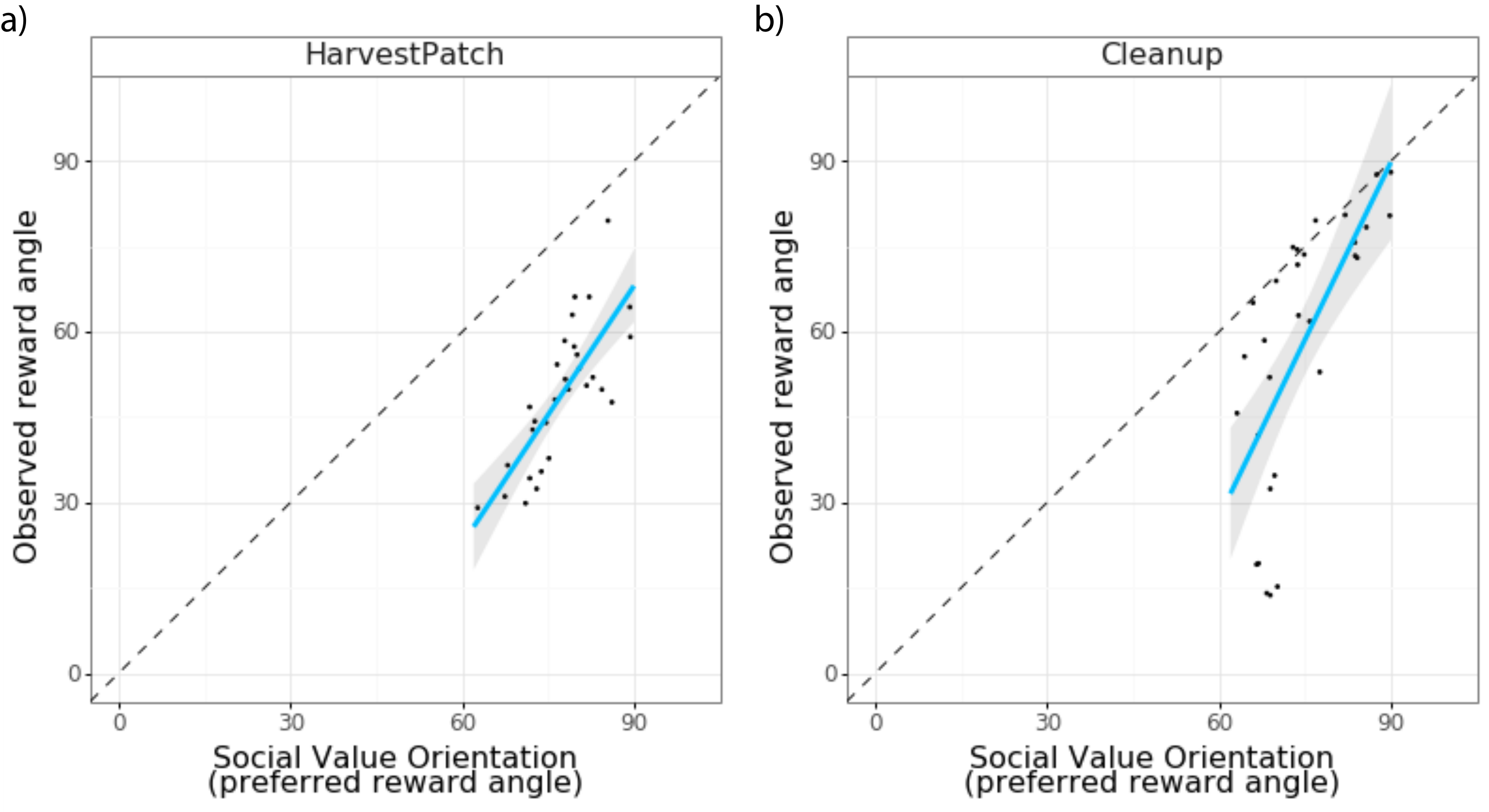}
    \caption{Correlation between target and observed reward angles among SVO agents in (a) HarvestPatch and (b) Cleanup. The higher an agent's SVO, the higher the reward angles it tended to observe.}
    \label{fig:agent_svo_reward_angles}
\end{figure}

How exactly does the distribution of SVO help diverse populations resolve these social dilemmas? We next evaluated the behavioral effects of SVO by examining a single, heterogeneous population within each task. We randomly selected two populations that achieved high equilibrium performance during training, parameterized with mean SVOs of $75\degree$ and standard deviations of $7.5\degree$. We gathered data from 100 episodes of play for both of these evaluation experiments, sampling 5 agents randomly for each episode. All regressions reported in this section are mixed error-component models, incorporating a random effect to account for the repeated sampling of individual agents. The accompanying figures depict average values per agent, with superimposed regression lines representing the fixed effect estimate ($\beta$) of SVO.

In our evaluation experiments, we observed a positive relationship between an agent's target reward angle and the group reward angles it tended to observe in HarvestPatch, $\beta = 1.51$, $p < 0.001$ (Figure \ref{fig:agent_svo_reward_angles}a). The effect of SVO on observed reward angle was similarly significant in Cleanup, $\beta = 2.08$, $p < 0.001$ (Figure \ref{fig:agent_svo_reward_angles}b). This reflects the association of higher agent SVO with the realization of more-prosocial distributions. In both environments, the estimated effect lies below the $45\degree$ line, indicating that agents acted somewhat more selfishly than their SVO would suggest.
% This selfish ``bias'' echoes the composition of $U_i$ from both environmental reward and the intrinsic reward for SVO (Eq. \ref{eqn:utility}).

\begin{figure}
    \centering
    \includegraphics[height=4.4cm]{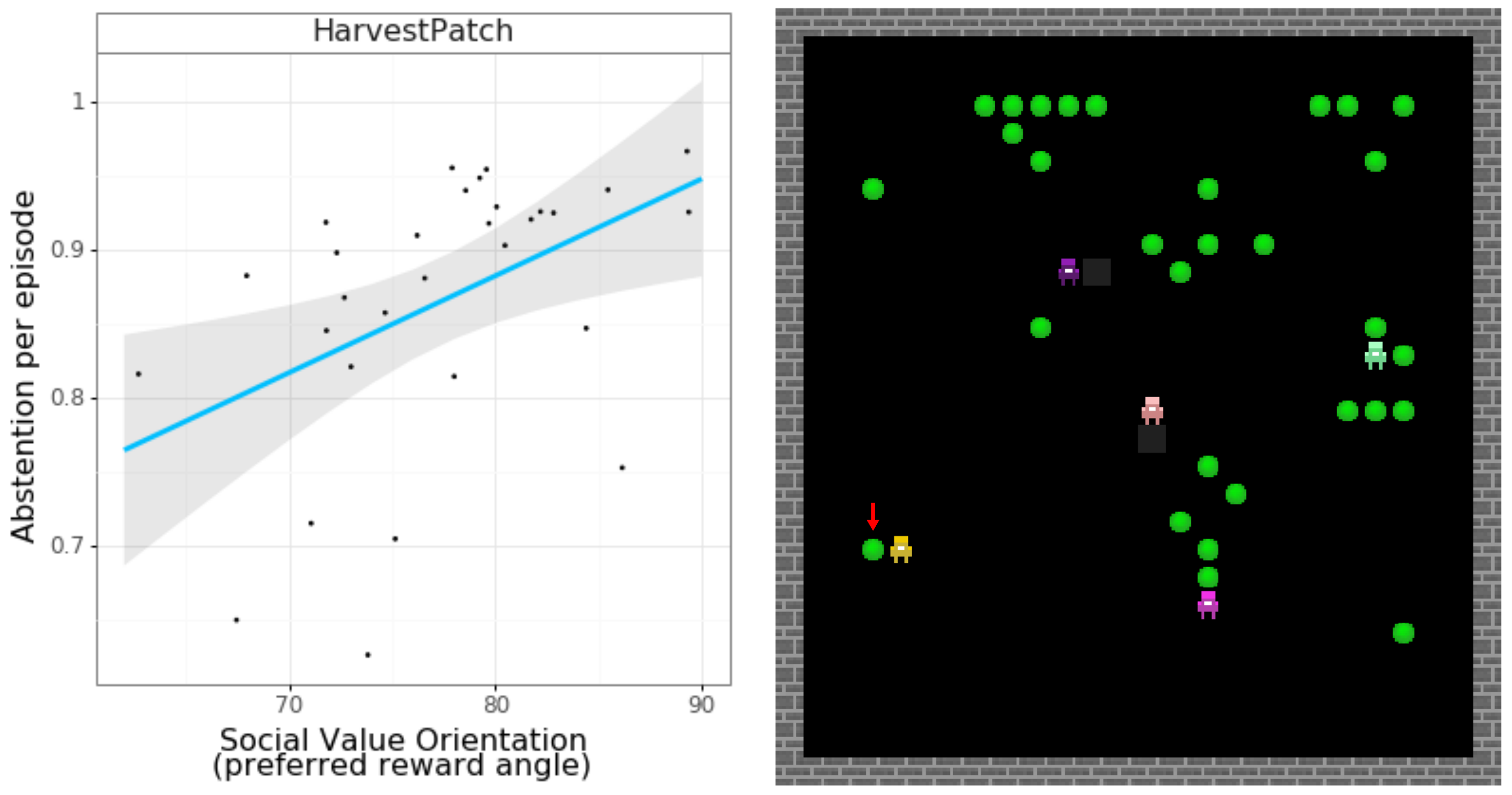}
    \caption{SVO and prosocial behavior in HarvestPatch. Agents with higher SVOs were significantly more likely to abstain from depleting local resource stocks. Here the yellow agent faces the choice of consuming an endangered apple for immediate reward or abstaining and traveling to a different patch.}
    \label{fig:patch_svo_prosocial}
\end{figure}

In HarvestPatch, an agent's prosociality can be estimated by measuring its abstention from consuming endangered apples. We calculated abstention as an episode-level metric incorporating the number of endangered apples an agent consumed and a normalization factor encoding at what points in the episode the endangered apples were consumed. An abstention score of 1 indicates that an agent did not eat a single endangered apple (or that it ate one or more endangered apples on the final step of the episode). An abstention score of 0, though not technically achievable, would indicate that an agent consumed one endangered apple from every apple patch in the environment on the first step of the episode. We observe a significant and positive relationship between an agent's SVO and its abstention, $\beta = 0.0065$, $p = 0.006$ (Figure \ref{fig:patch_svo_prosocial}).

\begin{figure}
    \centering
    \includegraphics[height=4.4cm]{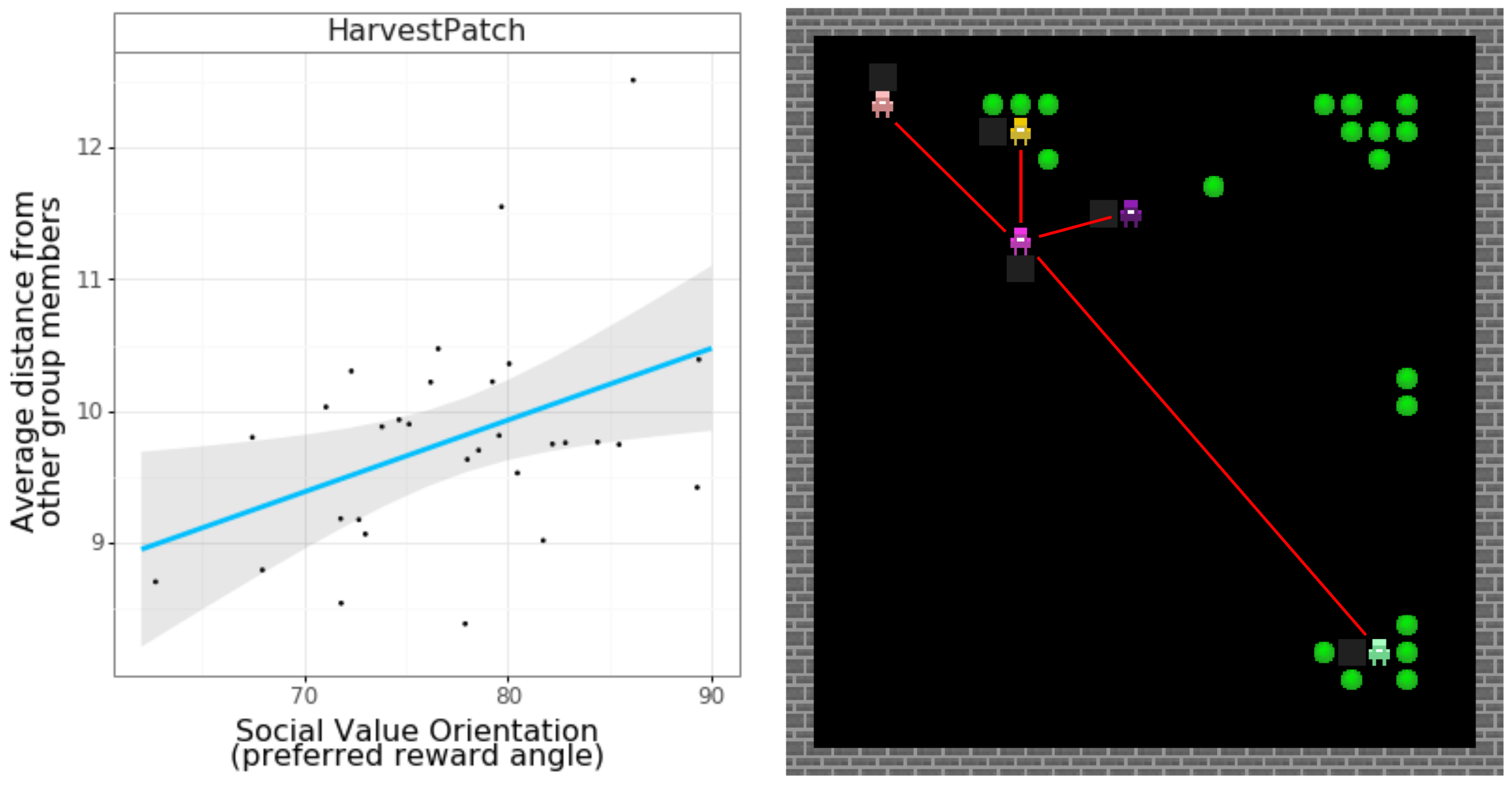}
    \caption{SVO and prosocial conventions in HarvestPatch. The higher an agent's SVO, the more distance it tended to maintain from other agents in its environment. Here the teal agent is maintaining a particularly high interagent distance, allowing it to sustainably harvest from a single patch.}
    \label{fig:patch_svo_norms}
\end{figure}

The structure of the HarvestPatch environment creates localized stock and flow components. Hosting too many agents in a single patch threatens to quickly deplete the local resource pool. Thus, one rule groups can use to maintain sustainability is for group members to harvest in separate patches, rather than harvesting together and sequentially destroying the environment's apple patches. We find that SVO correlated with distance maintained from other group members, $\beta = 0.005$, $p = 0.016$ (Figure \ref{fig:patch_svo_norms}). Consequently, groups with higher mean SVO established stronger conventions of interagent distance. This simple behavioral convention helped higher-SVO groups guard against environmental collapse.

\begin{figure}[b]
    \centering
    \includegraphics[height=4.4cm]{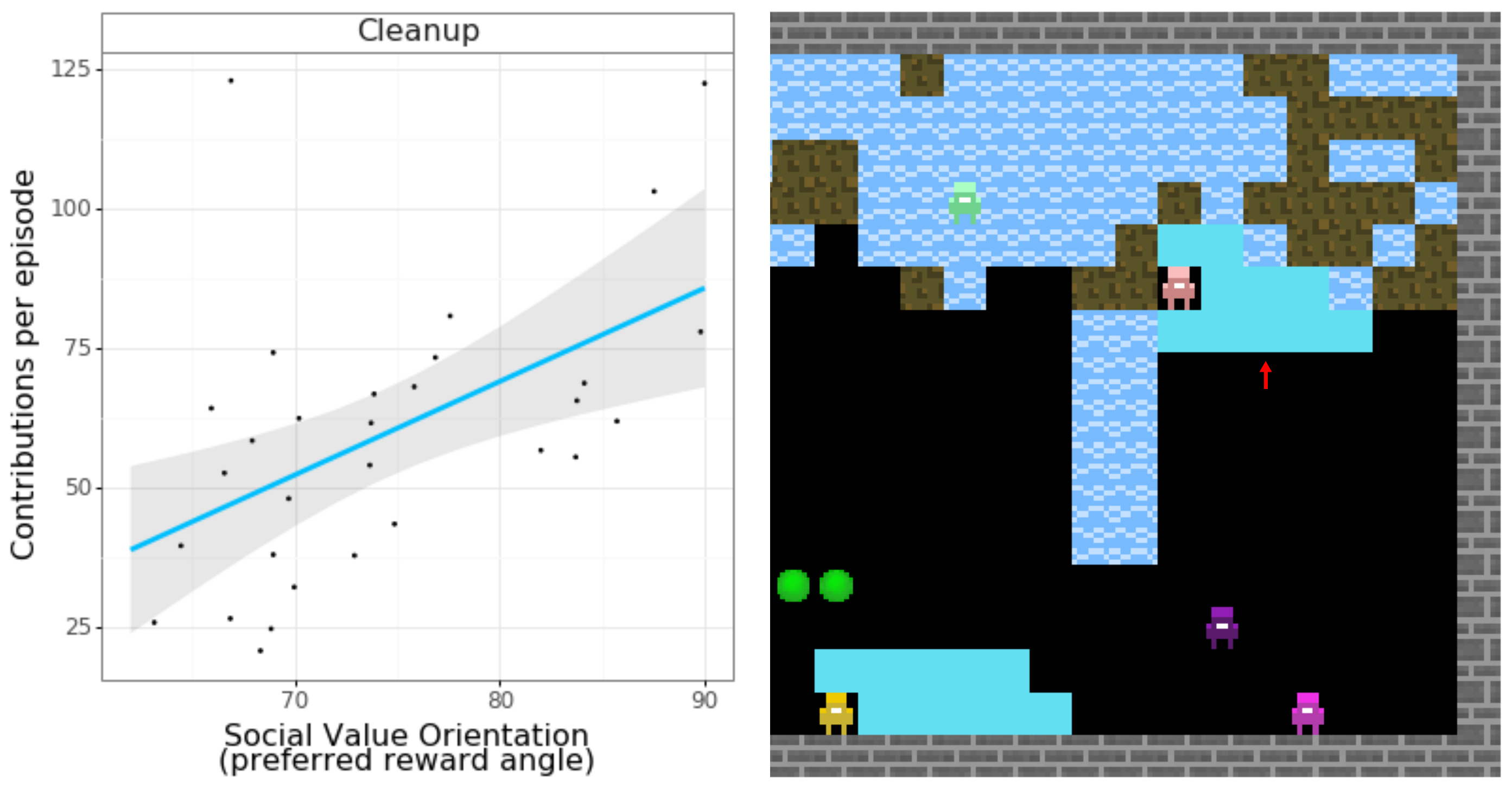}
    \caption{SVO and prosocial behavior in Cleanup. Agents with higher SVOs cleaned a significantly greater amount of pollution per episode than did peers with low SVO. Here the pink agent is actively cleaning two cells of pollution from the river. The yellow agent is using its cleaning action outside of the river, which does not affect its contribution score.}
    \label{fig:cleanup_svo_prosocial}
\end{figure}

In Cleanup, an agent's prosociality can be estimated by measuring the amount of pollution it cleans from the river. There was a significant and positive relationship between an agent's SVO and the amount of pollution it cleaned, $\beta = 1.68$, $p = 0.001$ (Figure \ref{fig:cleanup_svo_prosocial}). Agents with higher SVOs acted more prosocially by making greater contributions to the public good.

\begin{figure}
    \centering
    \includegraphics[height=4.4cm]{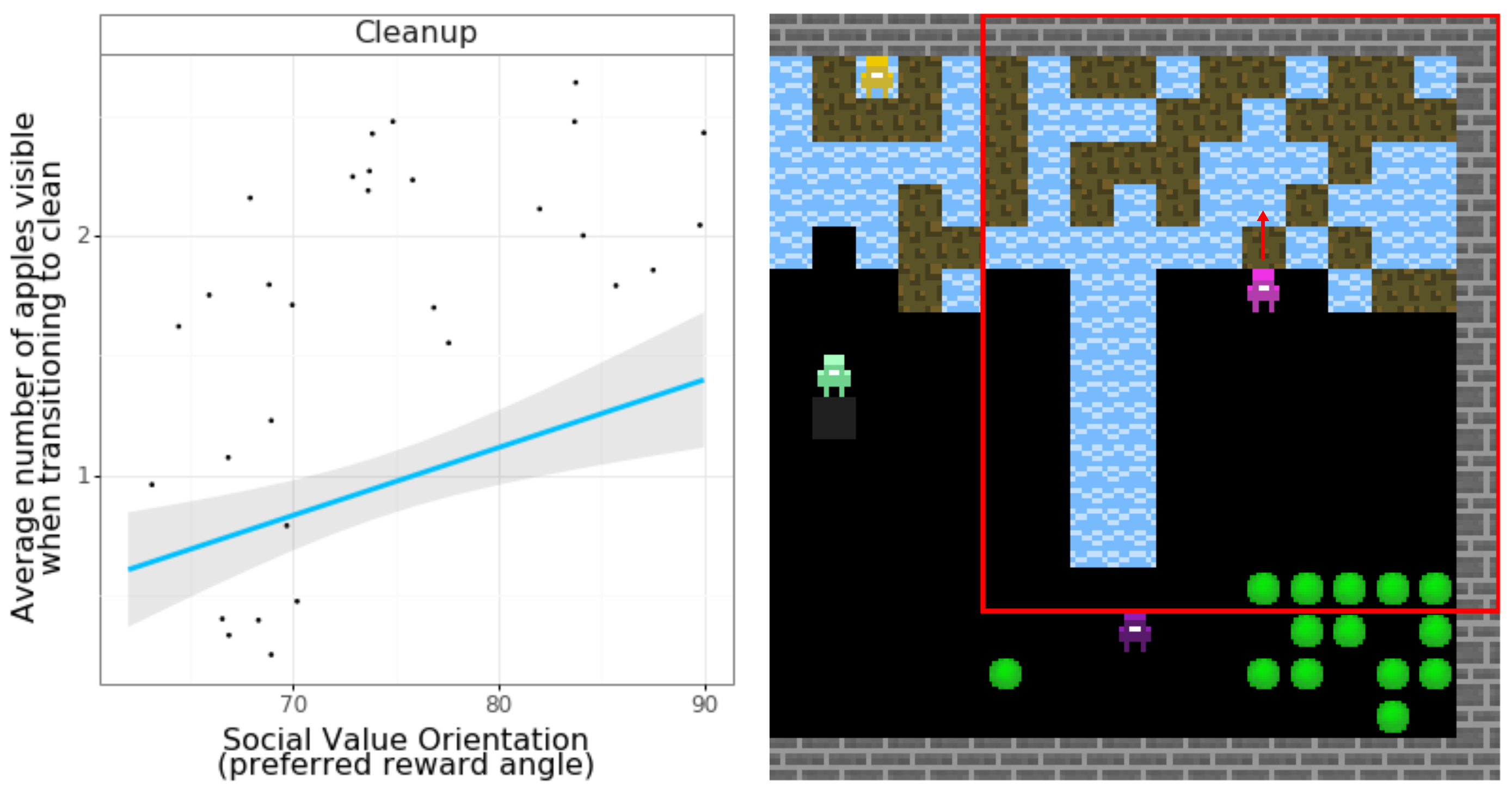}
    \caption{SVO and prosocial conventions in HarvestPatch. Agents with higher SVOs were significantly more likely to enter the river while there were unharvested apples within view. Here the magenta agent is transitioning to clean the river, even though it can observe multiple unharvested apples in the orchard.}
    \label{fig:cleanup_svo_norms}
\end{figure}

Finally, do SVO agents develop any sort of prosocial conventions in Cleanup to help maintain high levels of river cleanliness? In Cleanup, we examined one potential coordinating convention that we term behavioral preparedness: an inclination to transition from harvesting to cleaning even before the orchard is fully depleted. In Cleanup, groups that follow the short-term, individual-level incentive structure will respond primarily to the depletion of the orchard, rather than acting preventatively to ensure the public good is sustained over time. Groups that adopt welfare-maximizing strategies, on the other hand, will not wait for the orchard to be fully harvested to clean the river. We find a positive relationship between the average number of apples observable to agents at the times of their transitions to cleaning in each episode, $\beta = 0.028$, $p < 0.001$. The size and significance of this effect are not meaningfully affected by controlling for the number of times each agent transitioned to the river in a given episode, $\beta = 0.028$, $p < 0.001$ (Figure \ref{fig:cleanup_svo_norms}). In aggregate, this behavioral pattern helped high-SVO groups maintain higher levels of orchard regrowth over time.

\section{Discussion}

Recent research on pure-conflict and pure-cooperation reinforcement learning has highlighted the importance of developing robustness to diversity in opponent and partner policies \cite{carroll2019utility, jaderberg2019human, vinyals2019grandmaster}. We extend this argument to the mixed-motive setting, focusing in particular on the effects of heterogeneity in social preferences. Drawing from interdependence theory, we endow agents with Social Value Orientation (SVO), a flexible formulation for reward sharing among group members.

In the mixed-motive games considered here, homogeneous populations of pure altruists achieved high collective returns. However, these populations tended to produce hyper-specialized agents which reaped reward primarily from either intrinsic or extrinsic motivation, rather than both---a method of breaking the symmetry of the shared motivation structure. Thus, when equality-sensitive metrics are considered, populations with diverse distributions of SVO values were able to outperform homogeneous populations.

This pattern echoes the historic observation from interdependence theory that, if both players in a two-player matrix game adopt a ``maximize other's outcome'' transformation process, the resulting effective matrices often produce deficient group outcomes:

\medskip
\begin{adjustwidth}{0.35cm}{}
    ``It must be noted first that, in a number of matrices with a mutual interest in prosocial transformations, if one person acts according to such a transformation, the other is better off by acting according to his own \textit{given} outcomes than by adopting a similar transformation.'' \cite{kelley1978interpersonal}
\end{adjustwidth}
\medskip

\noindent This quote highlights a striking parallel between our findings and the predictions of interdependence theory. We believe this is indicative of a broader overlap in perspective and interests between multi-agent reinforcement learning and the social-behavioral sciences. Here we capitalize on this overlap, drawing inspiration from social psychology to formalize a general mechanism for reward sharing. Moving forward, SVO agents can be leveraged as a modeling tool for social psychology research \cite{morrison1999models}.

In this vein, \textit{group formation} is a topic important to both fields. It is well established among pyschologists that an individual's behavior is strongly guided by their ingroup---the group with which they psychologically identify \cite{de2010social}. However, the processes by which individuals form group identities are still being studied and investigated \cite{brewer1979group, turner2010social}. What sort of mechanisms transform and redefine self-interest to incorporate the interests of a broader group? This line of inquiry has potential linkages to the study of team and coalition formation in multi-agent research \cite{shenoy1979coalition}.

Our findings show that in multi-agent environments, heterogeneous distributions of SVO can generate high levels of population performance. A natural question follows from these results: how can we identify optimal SVO distributions for a given environment? Evolutionary approaches to reinforcement learning \cite{jaderberg2017population} could be applied to study the variation in optimal distributions of SVO across individual environments. We note that our results mirror findings from evolutionary biology that across-individual genetic diversity can produce group-wide benefits \cite{nonacs2007social}. We suspect that SVO agents can be leveraged \textit{in simulo} to study open questions concerning the emergence and adaptiveness of human altruism \cite{bowles2006group, mitteldorf2000population}.

The development of human-compatible agents still faces major challenges \cite{amershi2014power, amershi2019guidelines, ishowo2019behavioural}. In pure-common interest reinforcement learning, robustness to partner heterogeneity is seen as an important step toward human compatibility \cite{carroll2019utility}. The same holds true for mixed-motive contexts. Within ``hybrid systems'' containing humans and artificial agents \cite{christakis2019blueprint}, agents should be able to predict and respond to a range of potential partner behaviors. Social preferences are, of course, an important determinant of human behavior \cite{balliet2009social, kelley1978interpersonal}. Endowing agents with SVO is a promising path forward for training diverse agent populations, expanding the capacity of agents to adapt to human behavior, and fostering positive human-agent interdependence.